\documentclass[11pt]{article}
\usepackage{hyperref}
\pdfoutput=1
\begin{document}
\title{Spatially developing turbulent boundary layer on a flat plate}
\author{J. H. Lee, Y. S. Kwon, N. Hutchins and J. P. Monty \\
\\\vspace{6pt} Department of Mechanical Engineering, \\The University of Melbourne, Victoria, 3010 AUSTRALIA}

\maketitle

\begin{abstract}

This fluid dynamics video submitted to the Gallery of Fluid motion
shows a turbulent boundary layer developing under a 5 metre-long
flat plate towed through water. A stationary imaging system provides
a unique view of the developing boundary layer as it would form over
the hull of a ship or fuselage of an aircraft. The towed plate
permits visualisation of the zero-pressure-gradient turbulent
boundary layer as it develops from the trip to a high Reynolds
number state ($Re_\tau \approx 3000$). An evolving large-scale
coherent structure will appear almost stationary in this frame of
reference. The visualisations provide an unique view of the evolution
of fundamental processes in the boundary layer (such as interfacial
bulging, entrainment, vortical motions, etc.). In the more
traditional laboratory frame of reference, in which fluid passes
over a stationary body, it is difficult to observe the full
evolution and lifetime of turbulent coherent structures.  An
equivalent experiment in a wind/water-tunnel would require a camera
and laser that moves with the flow, effectively `chasing' eddies as
they advect downstream \cite{Longmire2011}.

\end{abstract}

\section*{Description of Experiment}
The experiment is conducted in a tow-tank of length 60 m, width 1.8
m and depth 2 m. The plate is towed by a fully automated carriage
capable of speeds up to $3$ ms$^{-1}$. The plate is 5.0 m long and
1.2 m wide with an elliptical leading edge. The boundary layer
formed on the bottom surface of the plate is tripped using a 1 mm
diameter trip wire. Fluorescein is injected through a spanwise
oriented slot located immediately downstream of the trip. The
developing boundary layer is then illuminated using a continuous 4 W
argon-ion laser which has been fanned into a streamwise /
wall-normal sheet. Images are acquired using a Redlake Y-3 classic
high-speed camera, with $1280 \times 1024$ pixel resolution. The
approximate field of view for all visualisations is $220 \times 180$
mm (streamwise $\times$ wall-normal).

Two flow visualisations are shown: one towed at a slower speed of
0.22 ms$^{-1}$ and the other at $0.9$ ms$^{-1}$. By the trailing
edge of the plate, the boundary layer develops to a friction
Reynolds number $Re_{\tau} \approx 1000$ and $Re_{\tau} \approx
3000$ respectively for the low and high speed experiments. Here
$Re_\tau = \delta U_{\tau}/\nu$, where $\delta$ is boundary layer
thickness, $U_{\tau}$ is wall-shear velocity and $\nu$ is the
kinematic viscosity. The camera frame-rate is 25 Hz for the low
speed, and 90 Hz for the high speed case.

By viewing the visualisations in slow motion, one can distinguish
the origins of the interfacial bulging that forms at the edge of the
boundary layer, leading to the well-known highly intermittent
fluctuation statistics in this region. One can also observe
eruptions of strongly vortical motions, originating close to the
wall, and growing beyond this interface. Connected with these events
is the entrainment of irrotational fluid from the freestream into
the boundary layer. These entrainment motions occasionally penetrate
deep into the boundary layer, transporting high momentum fluid from
the freestream to the near-wall region. By comparing the high and
low speed visualisations, the impact of Reynolds number on the
turbulent structure is clearly demonstrated. At the high Reynolds
number, the increased mixing reduces the contrast in the flow
visualisation. One can also observe smaller-scale motions and a more
complicated multi-scale interface separating the freestream from the
turbulence.

Ultimately this unique facility will be developed for high-speed
time-resolved particle image velocimetry (PIV) experiments to enable
statistical analysis of the evolving structure of turbulence. In the
meantime, these visualisations illustrate the complexity and beauty
of the fluid motion within wall-bounded turbulence.

\end{document}